\documentclass[12pt]{article}
\usepackage{amsmath, setspace}
\usepackage{graphicx}
\usepackage{lineno}

 \makeatletter       
 \renewcommand{\@biblabel}[1]{#1.}
 \makeatother
\begin{document}
\noindent
{\LARGE\textbf{Condition for the deflection of vertical cracks at dissimilar ice interfaces on Europa}}
\\
\\
{\large Daigo Shoji$^{1}$}\\
\\
1. Earth-Life Science Institute, Tokyo Institute of Technology, 2-12-1 Ookayama, Meguro-ku, Tokyo (shoji@elsi.jp)\\
\section*{Abstract}
The surface of Europa contains many quasi-circular morphologies called lenticulae. Although the formation mechanism of lenticulae is not understood, sill intrusion from the subsurface ocean is one promising hypothesis. However, it remains unclear how vertical cracks from the ocean deflect horizontally to allow sill intrusion in Europa. In this study, the critical stress intensity factor of Europan ice required for deflection was evaluated by considering crack theory at the interface between dissimilar materials and experimental results on ice. For deflection to occur at the interface between two dissimilar ices, the ratio of the critical stress intensity factor of the interface to that of the upper layer should be at most 0.45--0.5. This critical ratio may be attained if the interface is caused by brine-containing ice with a volume fraction of $>$30 ppt (3\%) and pure (no-brine) ice. Thus, a region with a temperature equal to the eutectic point (e.g., an area of approximately 240 K in the convective layer) is a candidate for the region in which the deflection occurs. 
\section{Introduction}
Europa, one of the four Galilean satellites orbiting Jupiter, is covered with an 80- to 170-km-thick H$_2$O layer [\textit{Anderson et al.}, 1998]. On the surface of Europa, circular or elliptical morphologies of a few to a few tens of kilometers in diameter have been observed [e.g., \textit{Rathbun et al.}, 1998; \textit{Collins et al.}, 2000; \textit{Culha and Manga}, 2016]. These structures, called lenticulae, are further classified into subcategories. Lenticulae with depressed surfaces are called pits, and convex lenticulae are defined as domes [\textit{Culha and Manga}, 2016]. Chaotic terrains are defined as circular uplifts with tilted and blocked preexisting crusts on their surfaces [\textit{Culha and Manga}, 2016]. 
On the basis of measurements of the induced magnetic field, Europa should have a subsurface ocean at a depth of a few to a few tens of kilometers [e.g., \textit{Kivelson et al.}, 2000]. Although many studies have considered the formation of lenticulae to be triggered by the interior ocean, the details of the formation mechanism are controversial, and several models have been suggested to explain lenticula formation [e.g., \textit{Rathbun et al.}, 1998; \textit{Collins et al.}, 2000; \textit{Sotin et al.}, 2002; \textit{Fagents}, 2003; \textit{Pappalardo and Barr}, 2004; \textit{Schmidt et al.}, 2011].  

Recently, \textit{Michaut and Manga} [2014] have suggested a detailed formation mechanism for pits, domes, and small chaotic terrains considering sill, which is intrusion of water in the horizontal direction. If a water sill intrudes into ice, the surface ice is depressed, leading to pit-type lenticulae. Then, after the sill freezes, the surface becomes convex because of the resulting volume expansion. This uplift can produce domes or small chaotic terrains. In this model, pits evolve into domes or small chaotic terrains. Thus, the sill model can explain the formation and evolution of different types of lenticulae with the same mechanism.

However, one question that remains unanswered well by the sill model is how a dyke deflects horizontally into the ice. Formation models of vertical dykes have been suggested. For example, the freezing of the ocean can induce a stress large enough to produce a crack at the base of the elastic layer [\textit{Manga and Wang}, 2007]. Because the time scale of crack propagation is much smaller than the Maxwell time (viscosity/rigidity) of ice, this crack can propagate to the subsurface ocean and to the surface. If the strength of the ice is reduced sufficiently to approximately 10$^4$--10$^5$ Pa [e.g., \textit{Dempsey et al.}, 1999; \textit{Lee et al.}, 2005], a tidal stress of approximately 10$^5$ Pa [\textit{Wahr et al.}, 2009] may also generate a crack at the bottom of the ice layer [\textit{Crawford and Stevenson}, 1988]. Nonsynchronous rotation also produces large tensile stress [\textit{Wahr et al.}, 2009], which may induce cracking. Detailed numerical simulations have revealed that, once a crack is formed, a stress of a few MPa, which is larger than the strength of ice, is induced at the crack tip [\textit{Craft et al.}, 2016]. Once the vertical crack comes into contact with the ocean, water fills the crack up to the depth at which the buoyancy becomes neutral. 

Thus, it is probable that vertical dykes of water are formed in Europa's ice layer.  However, it cannot be said that the vertical crack necessarily deflects horizontally at the neutral buoyancy level. In terrestrial rock, deflections of dykes have been observed at heights different from the neutral buoyancy level of magma [\textit{Kavanagh et al.}, 2006]. \textit{Craft et al.} [2016] performed numerical simulations on dyke propagation to determine the condition for deflection. Although they found that the effect of tidal bulges may cause the dyke direction to change, the deflection of vertical cracks is still challenging to understand [\textit{Craft et al.}, 2016].    

In this work, to link the formation of vertical dykes [\textit{Crawford and Stevenson}, 1988; \textit{Manga and Wang}, 2007] and the evolution of water sills [\textit{Michaut and Manga}, 2014; \textit{Manga and Michaut}, 2017], the deflection of cracks at the interface between materials with dissimilar Young's moduli is considered. In the case of terrestrial studies, many sills of magma have been observed at interfaces between materials of different stiffness [\textit{Kavanagh et al.}, 2006]. Thus, although several causes, such as the surface compressive stress [\textit{Menand et al.}, 2010], have been suggested, a discontinuous interface is one of the most promising hypotheses regarding the formation of sills [e.g., \textit{Kavanagh et al.}, 2006; \textit{Gudmundsson}, 2011]. The effect of the different Young's moduli on the deflection has been evaluated by \textit{Craft et al.} [2016], and they found that this effect was very small. However, whether a crack deflects strongly depends on the toughness of the ice [\textit{He and Hutchinson}, 1989; \textit{He et al.}, 1994], which is related to the critical stress intensity factor as well as the Young's modulus. The critical stress intensity factor is strongly dependent on the material and conditions, and should thus be determined experimentally.  Currently, the material toughness of Europan ice is unknown. Here, a condition for crack deflection was determined by applying critical stress intensity factors obtained from laboratory experiments on ice.

\section{Condition for deflection}
The dissimilarity of two materials can be parameterized by Dundurs' parameters $\alpha$ and $\beta$ [\textit{He and Hutchinson}, 1989; \textit{He et al.}, 1994], which are given by 

\begin{equation}
\alpha=\frac{\bar{E}_1-\bar{E}_2}{\bar{E}_1+\bar{E}_2}
\label{alpha}
\end{equation}
and
  
\begin{equation}
\beta=\frac{1}{2}\frac{\mu_1(1-2\nu_2)-\mu_2(1-2\nu_1)}{\mu_1(1-\nu_2)+\mu_2(1-\nu_1)},
\label{beta}
\end{equation}
where $\nu$ and $\mu$ are the Poisson's ratio and shear modulus, respectively [\textit{He and Hutchinson}, 1989; \textit{He et al.}, 1994], and the subscripts 1 and 2 represent the two materials (materials 1 and 2 in Fig.~\ref{pene_def}). Additionally, $\bar{E}=E/(1-\nu^2)$, where $E$ is the Young's modulus. As shown below, in the case of Europa, materials 1 and 2 can be regarded as the upper and lower dissimilar ice layers, respectively (e.g., warm convective ice under a cold stagnant lid or brine-containing ice under pure ice). If the Poisson's ratio is fixed, both $\alpha$ and $\beta$ depend only on the Young's modulus because the shear modulus can be represented as $\mu=E/2(1+\nu)$. Here, $\nu_1$ and $\nu_2$ are assumed to be 0.33 for the sake of simplicity. Measurements of terrestrial sea ice have shown that the Poisson's ratio can reach up to approximately 0.42 [\textit{Timco and Weeks}, 2010]. However, in this range, the effect of varying the Poisson's ration is small. The Young's modulus $E_1$ of the upper layer scales as $E_1=\gamma E_2$, where $\gamma$ is the ratio of $E_1$ to $E_2$. Because of this scaling, $E_1$ and $E_2$ can be eliminated from Eqs.~(\ref{alpha}) and (\ref{beta}), allowing $\alpha$ and $\beta$ to be calculated from $\gamma$.

When a crack propagating in material 2 penetrates material 1 (Fig.~\ref{pene_def}), propagation after penetration is caused by mode I fracture (crack opening by tensile stress). Thus, the strain energy release rate $G_p$ of crack propagation in material 1 is given by [\textit{He and Hutchinson}, 1989; \textit{He et al.}, 1994]

\begin{equation}
G_p=\frac{1-\nu_1}{2\mu_1}(K_{\mathrm{I}})_1^2,
\label{pene}
\end{equation}
where $(K_\mathrm{I})_1$ is the stress intensity factor of mode I cracking in material 1. In contrast to the penetration of the crack into material 1, propagation along the interface between the two materials is caused by the mixing of mode I and II fracture (sliding crack) (Fig.~\ref{pene_def}). The strain energy release rate at the interface $G_d$ is given by

\begin{equation}
G_d=\frac{1}{4}\left(\frac{1-\nu_1}{\mu_1}+\frac{1-\nu_2}{\mu_2}\right)\frac{(K_\mathrm{I})_i^2+(K_{\mathrm{II}})_i^2}{\cosh^2(\pi\epsilon)},
\label{def}
\end{equation}
where $(K_{\mathrm{I}})_i$ and $(K_{\mathrm{II}})_i$ are the stress intensity factors of mode I and II cracking at the interface, respectively [\textit{He and Hutchinson}, 1989]. $\epsilon$ is defined as

\begin{equation}
\epsilon=\frac{1}{2\pi}\ln\left(\frac{1-\beta}{1+\beta}\right).
\label{epsilon}
\end{equation}

At the critical stress intensity factor, where  $(K_\mathrm{I})_1=(K_{\mathrm{I}c})_1$, $(K_\mathrm{I})_i=(K_{\mathrm{I}c})_i$, and $(K_{\mathrm{II}})_i=(K_{\mathrm{II}c})_i$, $G_p$ and $G_d$ can be regarded as the material toughnesses $G_{pc}$ and $G_{dc}$ for penetration and deflection, respectively. It has been revealed that, for the deflection of the crack to occur, the ratio of the two toughnesses $G_{dc}/G_{pc}$ must be smaller than the critical value $(G_d/G_p)_{\mathrm{def}}$ [\textit{He and Hutchinson}, 1989; \textit{He et al.}, 1994]; otherwise, the crack penetrates the interface instead of deflecting. $(G_d/G_p)_{\mathrm{def}}$ is given by

\begin{equation}
\left(\frac{G_d}{G_p}\right)_{\mathrm{def}}=\frac{|d|^2}{c(1-\alpha)},
\label{ratio}
\end{equation}
where $c$ and $d$ are real and complex functions, respectively, of $\alpha$ and $\beta$ [\textit{He et al.}, 1994]. Although \textit{He et al.} [1994] took into account the effect of the residual stress, it is ignored in this work for the sake of simplicity. To determine $c$ and $d$, integral equations must be solved. However, \textit{He et al.} [1994] gave values of $c$ and $d$ for different values of $\alpha$ at $\beta=0$ (see Tables 1 and 2 in \textit{He et al.} [1994]). Because the two functions depend much more strongly on $\alpha$ than on $\beta$, the values given by \textit{He et al.} [1994] were used for the present evaluation of Europan ice.  

From Eqs.~(\ref{alpha})--(\ref{ratio}) and the condition for deflection $G_{dc}/G_{pc}<(G_d/G_p)_{\mathrm{def}}$, the critical stress intensity factor ratio to induce crack deflection must satisfy

\begin{equation}
\frac{(K_{\mathrm{I}c})_i^2+(K_{\mathrm{II}c})_i^2}{(K_{\mathrm{I}c})_1^2}<\frac{1}{F}\frac{|d|^2}{c(1-\alpha)},
\label{critical}
\end{equation}
where 

\begin{equation}
F=\frac{\mu_1}{2(1-\nu_1)\cosh^2(\pi\epsilon)}\left(\frac{1-\nu_1}{\mu_1}+\frac{1-\nu_2}{\mu_2}\right).
\label{factor}
\end{equation}

Although experiments on the critical stress intensity factor of ice under mixed-mode fracture is not sufficient, \textit{Shen and Lin} [1988] have shown that $K_{\mathrm{I}c}^2+K_{\mathrm{II}c}^2$ under mixed-mode fracture is almost the same as $K_{\mathrm{I}c}^2$ under pure mode I cracking. Thus, with this approximation of the critical intensity factor in mixed-mode cracking to that in pure mode I cracking for simplicity, the condition for deflection is given by

\begin{equation}
\frac{(K_{\mathrm{I}c})_i}{(K_{\mathrm{I}c})_{1}}<\left(\frac{K_d}{K_p}\right)_{\mathrm{def}},
\label{critical2}
\end{equation}
where 

\begin{equation}
\left(\frac{K_d}{K_p}\right)_{\mathrm{def}}=\sqrt{\frac{|d|^2}{Fc(1-\alpha)}}.
\label{critical_k}
\end{equation}
In the case where $(K_{\mathrm{I}c})_i/(K_{\mathrm{I}c})_{1}>(K_d/K_p)_{\mathrm{def}}$, the crack penetrates material 1. Hereafter, $(K_{\mathrm{I}c})_i$ and $(K_{\mathrm{I}c})_{1}$ are referred to as the critical intensity factors in pure mode I fracture.

\begin{figure}[t]
\centering
\includegraphics[width=15cm] {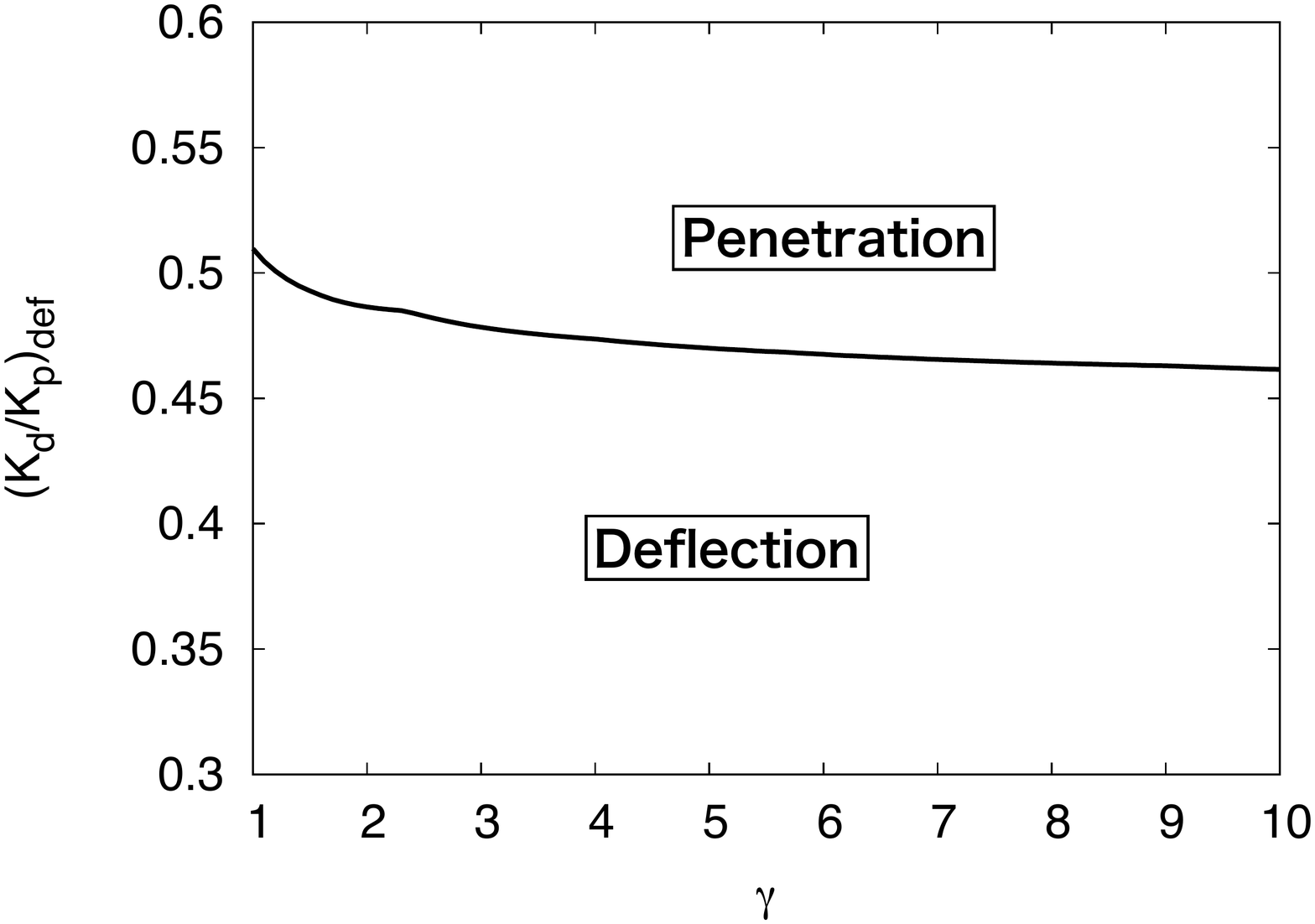}
\caption{$(K_d/K_p)_{\mathrm{def}}$ as a function of $\gamma=E_1/E_2$. If $(K_{\mathrm{I}c})_i/(K_{\mathrm{I}c})_{1}$ is smaller than $(K_d/K_p)_{\mathrm{def}}$, the crack deflects and propagates along the interface.}
\label{crit_k}
\end{figure}

Fig.~\ref{crit_k} shows $(K_d/K_p)_{\mathrm{def}}$ as a function of $\gamma=E_1/E_2$. Although the variation is small, $(K_d/K_p)_{\mathrm{def}}$ decreases slightly with increasing $\gamma$. Large $\gamma$ means that the difference between the Young's moduli of the two materials is large, which results in an increase in $\alpha$ and works to increase $(K_d/K_p)_{\mathrm{def}}$ [\textit{He et al.}, 1994]. However, $F$ also increases with increasing $\gamma$, which reduces $(K_d/K_p)_{\mathrm{def}}$ at large $\gamma$. The values of $(K_d/K_p)_{\mathrm{def}}$ show that if the value of $(K_{\mathrm{I}c})_i/(K_{\mathrm{I}c})_{1}$ for Europan ice is as low as 0.45--0.5, a vertical crack can deflect at the interface. This result changes little even if the Poisson's ratio of Europa is considered to be approximately 0.42. Because the dependence of $(K_d/K_p)_{\mathrm{def}}$ on the Young's modulus difference ($\gamma$) is not so strong, the magnitude of the critical stress intensity factor is more important to estimate crack deflection in Europa. 

\section{Mechanism to generate interface}
From Fig.~\ref{crit_k}, for a vertical crack to deflect horizontally, $(K_{\mathrm{I}c})_i/(K_{\mathrm{I}c})_{1}$ should be reduced to $<$0.5. The critical stress intensity factor of the ice is dependent on several conditions, such as the temperature or the brine volume fraction of the ice. 
Fig.~\ref{model}(a) shows a schematic view of Europa's ice layer. Because the surface temperature of Europa is approximately 100 K, the viscosity varies greatly through the ice layer, which causes stagnant lid-type convection, and the ice layer is mainly separated into a convective layer and a stagnant lid [e.g., \textit{Han and Showman}, 2011]. The convective layer is further separated into a well-mixed phase with an adiabatic temperature gradient and a conductive phase in which heat is transferred by conduction. The mean temperature of the well-mixed part and the temperature of the bottom of the stagnant lid are approximately 250 and 197 K, respectively [\textit{Nimmo and Manga}, 2002; \textit{Pappalardo et al.}, 1998]. In this section, mechanisms to attain a small critical stress intensity factor ratio are discussed on the basis of this interior structure. 

\subsection{Porosity}
The surface porosity of Europa can be a few tens of percent [\textit{Lee et al}., 2005] within a depth of approximately 3 km [\textit{Nimmo et al.}, 2003]. Increasing porosity reduces the critical stress intensity factor. Although the experimental results are scattered, the relationship between the critical stress intensity factor $K_{\mathrm{I}c}$ and the porosity $p$ can be represented linearly as $K_{\mathrm{I}c}(p)=K_{\mathrm{I}c}(1-p)$ for $p<0.4$ [\textit{Schulson and Duval}, 2009]. Thus, even though a discontinuous interface in the porosity exists at a depth of approximately 3 km, this interface works to increase $(K_{\mathrm{I}c})_i/(K_{\mathrm{I}c})_{1}$ by reducing $(K_{\mathrm{I}c})_1$, and crack deflection becomes less probable.

\subsection{Temperature}
Several experiments have shown that the temperature dependence of the critical stress intensity factor is weak [\textit{Schulson and Duval}, 2009]. However, \textit{Litwin et al.} [2012] have revealed that $K_{\mathrm{I}c}$ decreases if the temperature is larger than approximately 250 K. This temperature matches well with the mean temperature of the well-mixed phase in the convective layer (Fig.~\ref{model}(a)). Thus, the interface between the conductive and well-mixed phases may be suitable for deflection.
However, the rate of decrease in the critical stress intensity with increasing temperature is approximately 30\% [\textit{Litwin et al.}, 2012]. Thus, $(K_{\mathrm{I}c})_i/(K_{\mathrm{I}c})_{1}$ is approximately 0.7 under the assumption that $(K_{\mathrm{I}c})_i$ is similar to $K_{\mathrm{I}c}$ of the well-mixed phase, which does not meet the condition for the deflection of the crack ($(K_{\mathrm{I}c})_i/(K_{\mathrm{I}c})_{1}<0.5$). 

\subsection{Grain size}
Although the grain size dependence of the critical stress intensity factor is not so clear [\textit{Schulson and Duval}, 2009], the relationship between the critical stress intensity factor and the grain size has been experimentally obtained as [\textit{Nixon and Schulson} 1988]

\begin{equation}
K_{\mathrm{I}c}\;[\mathrm{kPa\;m}^{1/2}]=58.3+42.4d^{-1/2},
\label{grainsize}
\end{equation}
where $d$ is the grain size in millimeters. Thus, ice with a large grain size results in a small critical stress intensity factor. 

\textit{Barr and McKinnon} [2007] have calculated the distribution of the convective ice layer using the finite element method, and revealed that, because of recrystallization, the convective layer has larger grains than the stagnant lid. Thus, the boundary between the stagnant lid and the convective layer may be the interface for deflection. However, considering Eq.~(\ref{grainsize}) and the minimum grain size of approximately 1 mm for icy satellites, $(K_{\mathrm{I}c})_i/(K_{\mathrm{I}c})_{1}$ cannot be as low as 0.5 regardless of the grain size in the convective layer. In addition, for convection to continue, the grain size should be less than a few tens of millimeters because the viscosity increases with increasing grain size [\textit{Barr and McKinnon}, 2007]. \textit{Han and Showman} [2011] considered the effect of tidal heat. In their calculations, Europan ice can maintain convection at a grain size of 1--10 mm. When the grain sizes in the lid and convective layer are respectively 1 and 10 mm, $(K_{\mathrm{I}c})_i/(K_{\mathrm{I}c})_{1}$ is approximately 0.7. Thus, as long as Europan ice convects for a long time, the effect of the grain size is not sufficient to cause the deflection of crack.
 
\subsection{Brine}
Measurements of infrared spectra have indicated that hydrated magnesium sulfates and sodium carbonates are present on the surface of Europa [e.g., \textit{Carlson et al}., 1999; \textit{McCord et al.}, 1998; \textit{Orlando et al.}, 2005; \textit{Dalton}, 2007]. Sodium has also been observed in the Europan atmosphere [\textit{Brown and Hill}, 1996; \textit{Johnson}, 2000; \textit{Brown}, 2001]. Although the origin (endogenic or exogenic) of the chemical composition, especially sulfur, requires further study, it is highly probable that the Europan ocean is rich with chemicals such as Na$_2$SO$_4$, MgSO$_4$, and NaCl [e.g., \textit{Zolotov and Shock}, 2003; \textit{Brown and Hand}, 2013; \textit{Hand and Carlson}, 2015; \textit{Marison}, 2002]. Studies on the interaction between ice and the ocean have indicated that chemical composition in the ocean can be trapped in the ice and transported upward by convection [\textit{Zolotov et al.}, 2004; \textit{Peddinti and McNamara}, 2015]. Thus, brine is generated in the ice layer when the temperature is larger than the eutectic point.

Although the amount of available data is not sufficient to draw strong conclusions and the results are scattered, when ice contains brine, the critical stress intensity factor tends to decrease. \textit{Timco and Frederking} [1983] have shown that $K_{\mathrm{I}c}$ with a brine volume fraction of approximately 50 ppt can decrease to  approximately 80 kPa~m$^{1/2}$ from approximately 160 kPa m$^{1/2}$ in the case with no brine. Experiments performed by \textit{Urabe and Yoshitake} [1981] with a higher brine fraction demonstrated that it is probable that $K_{\mathrm{I}c}$ is reduced to as low as approximately 30--40 kPa~m$^{1/2}$ at a brine volume fraction of approximately 130 ppt [\textit{Timco}, 1985]. Because $K_{\mathrm{I}c}$ without brine is approximately 160 kPa m$^{1/2}$ [\textit{Timco and Frederking}, 1983], the interface between ice layers with and without brine may satisfy the condition for deflection. 

\begin{figure}[t]
\centering
\includegraphics[width=15cm] {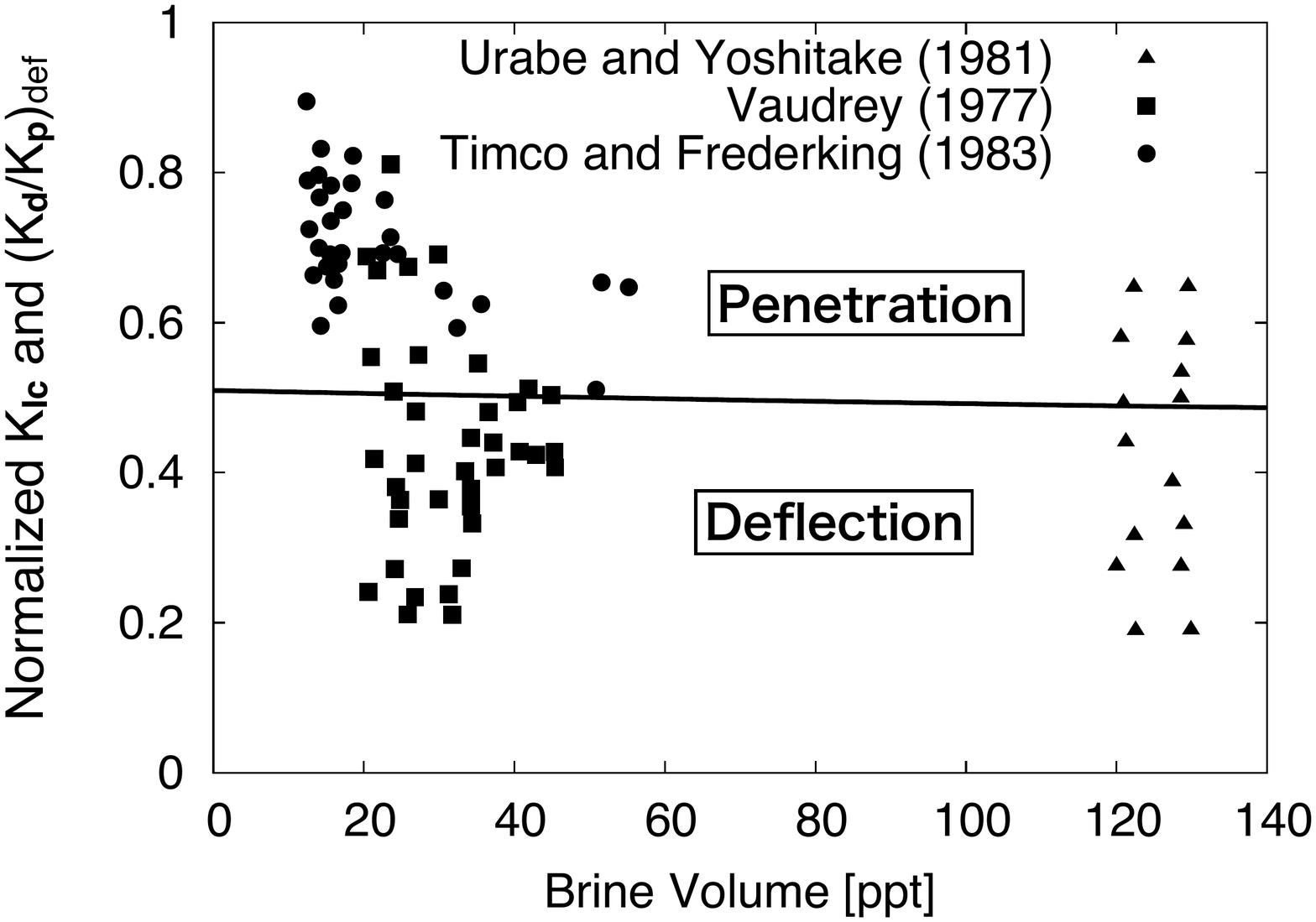}
\caption{$(K_d/K_p)_{\mathrm{def}}$ and $K_{\mathrm{I}c}$ plotted against the brine volume fraction. For comparison, $K_{\mathrm{I}c}$ is normalized by 160 kPa m$^{1/2}$, which can be regarded as the critical stress intensity factor $(K_{\mathrm{I}c})_1$ of pure ice. The experimental values were obtained from \textit{Timco} [1985].}
\label{brine}
\end{figure}

The Young's modulus as a function of brine volume $v_b$ is given by [\textit{Timco and Weeks}, 2010]

\begin{equation}
E\;[\mathrm{GPa}]=10-0.0351v_b\;[\mathrm{ppt}].
\label{young}
\end{equation}
From Eqs.~(\ref{alpha})--(\ref{critical_k}) and (\ref{young}), a relationship between $(K_d/K_p)_{\mathrm{def}}$ and $v_b$ can be calculated. Fig.~\ref{brine} shows experimentally obtained values of $K_{\mathrm{I}c}$ [\textit{Timco}, 1985] and $(K_d/K_p)_{\mathrm{def}}$ at different brine volume fractions. $E_1$ was assumed to be 10 GPa (no brine). The experimental values of $K_{\mathrm{I}c}$ are normalized by 160 kPa m$^{1/2}$, which is consistent with the critical intensity without brine [\textit{Timco and Frederking}, 1983] and can be regarded as $(K_{\mathrm{I}c})_1$. Although the experimental results are scattered, on the basis of the assumption that $(K_{\mathrm{I}c})_i$ is not so different from $K_{\mathrm{I}c}$ of briny ice, $(K_{\mathrm{I}c})_i/(K_{\mathrm{I}c})_1$ may be smaller than $(K_d/K_p)_{\mathrm{def}}$ at $v_b>30$ ppt. Some experiments have suggested that the $K_{\mathrm{I}c}$ value of fresh ice is 120--150 kPa m$^{1/2}$ [\textit{Schulson and Duval}, 2009; \textit{Litwin et al.}, 2012]. However, even in the case of smaller $(K_{\mathrm{I}c})_1$, $(K_{\mathrm{I}c})_i/(K_{\mathrm{I}c})_1$ can be lower than the critical value at brine volume fractions of more than approximately 30--40 ppt. Thus, it is probable that the crack in Europa deflects along the interface if the amount of brine generated by eutectic ice exceeds 30 ppt (3\%). Although the detailed composition and salinity of Europan ice have not been well characterized, in the case of terrestrial sea ice, the brine volume fraction can reach up to 20\% [\textit{Thomas and Dieckmann}, 2002; \textit{Galley et al.}, 2015]. Thus, a brine volume fraction of 3--4\% may be generated in Europa.

\begin{figure}[t]
\centering
\includegraphics[width=13cm] {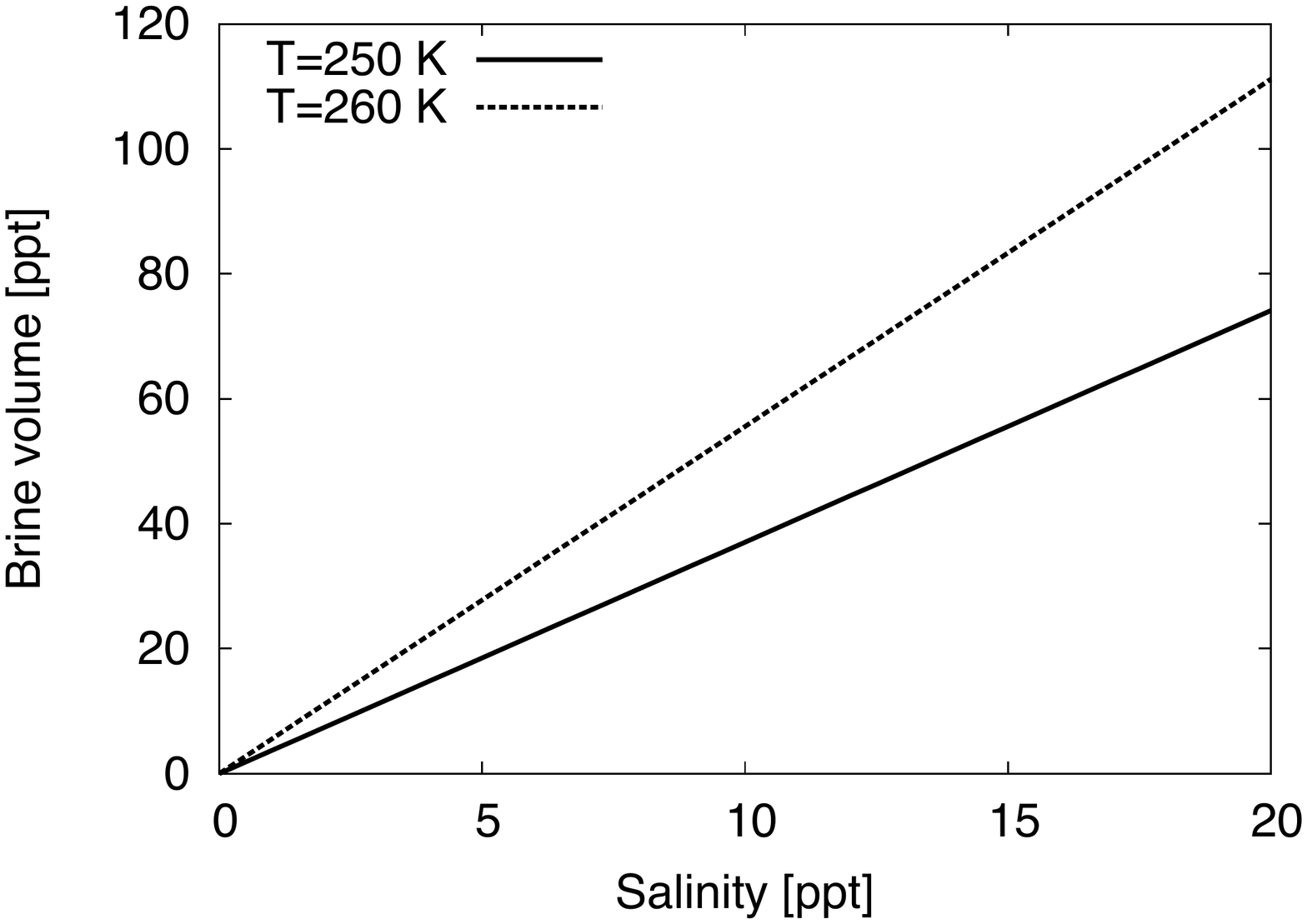}
\caption{Brine volume fraction of ice as a function of salinity.}
\label{salinity_brine}
\end{figure}

The relationship between the salinity of ice $S_i$ [ppt] and brine volume fraction $v_b$ can be given by [\textit{Cox and Weeks}, 1983]

\begin{equation}
v_b\;\mathrm{[ppt]}=\frac{\rho_{bulk}S_i}{F_1(T)}
\label{salinity}
\end{equation}
where $\rho_{bulk}$ [kg~m$^{-3}$] is the bulk density of brine-containing ice and at $T\geq250$ K,

\begin{equation}
F_1(T)=-4.732-22.45(T-273)-0.6397(T-273)^2-0.01074(T-273)^3. 
\label{f}
\end{equation}
Fig.~\ref{salinity_brine} shows $v_b$ as a function of $S_i$ under the assumption that $\rho_{bulk}=1126$ kg~m$^{-3}$, which is the density of MgSO$_4$--Na$_2$SO$_4$--H$_2$O eutectic ice [\textit{Williams and Greeley}, 1998]. Although the composition of Europan ice may be significantly different from that of terrestrial sea ice, if the salinity of the ice is more than 10 ppt (1\%), a brine volume of approximately 40 ppt can be generated. $\rho_{bulk}$ changes with the chemical composition of the ice. However, within a reasonable range  for Europan ice (1000--1200 kg~m$^{-3}$), a sufficient brine volume fraction can be generated. 

\section{Deflection mechanisms based on a brine interface}
\begin{figure}[t]
\centering
\includegraphics[width=17cm] {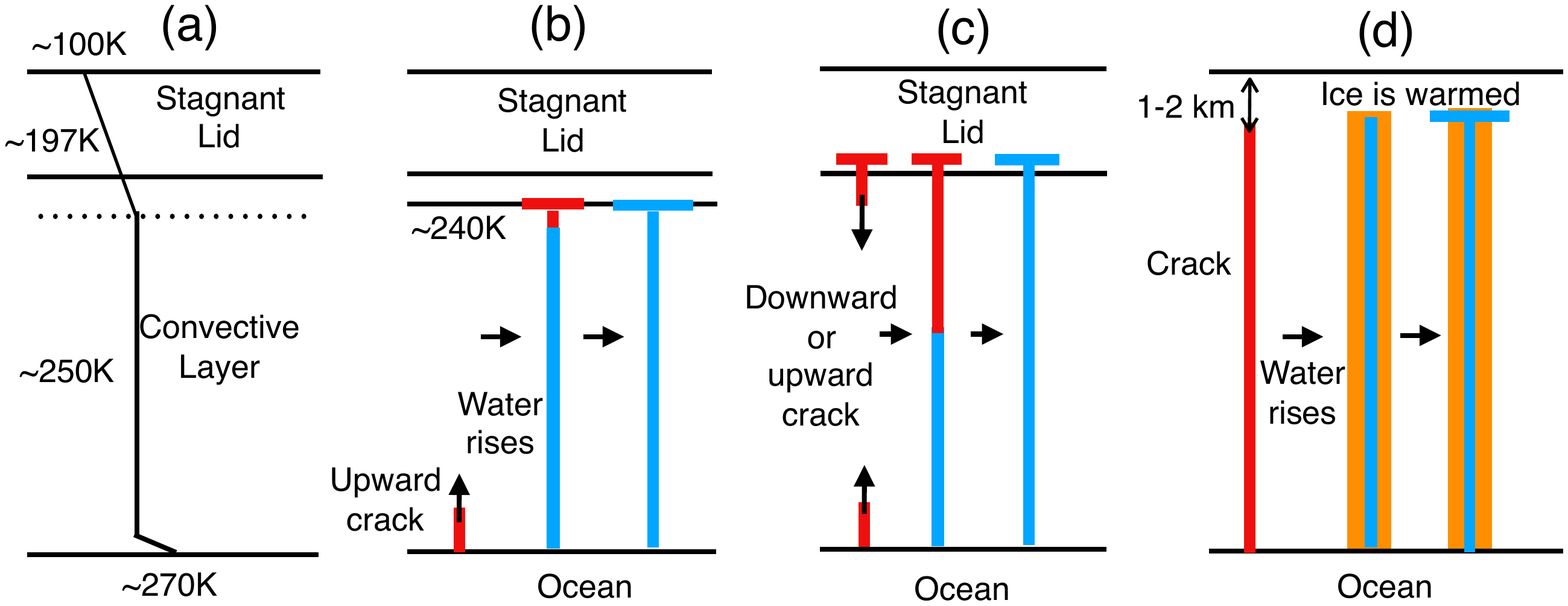}
\caption{(a) Schematic view of Europa's convective ice layer and (b)--(d) three mechanisms that may induce deflection at a brine interface. (b) The crack propagates from the bottom of the ice and deflects when the temperature is approximately 240 K, which is the eutectic point of Europan ice. (c) The stagnant lid contains components with low eutectic points, and the brine interface is in the stagnant lid. The crack is  produced by volume expansion at the base of the lid or tidal stress at the bottom of the convective layer. (d) The crack propagates up to a depth of 1--2 km and water fills the crack, which warms the surrounding eutectic ice. An interface is generated at the top of the dyke, and deflection occurs by tidal stress.}
\label{model}
\end{figure}

Of the mechanisms described above, the existence of an interface between layers of ice with and without brine is a strong candidate for the deflection. Combining the crack formation models and the interior structure of Europa's ice layer, three types of deflection are proposed, as shown in Fig.~\ref{model}. In the first mechanism, the well-mixed phase of the convective layer contains brine, and a vertical crack generated at the bottom of the ice propagates upward. Then, the vertical crack is deflected at the interface between the brine-containing and no-brine ice regions in the convective layer (Fig.~\ref{model}(b)). As demonstrated by numerical simulations conducted by \textit{Zolotov et al.} [2004], brine can be contained in ice when the ice temperature is greater than 240 K. Thus, deflection may occur in the conductive part of the convective layer. Because this temperature is too high for ice to show elastic behavior, downward cracks produced by volume expansion [\textit{Manga and Wang}, 2007] cannot be generated. One problem is that, although sufficient stress is required to induce crack formation, the magnitude of the tidal stress is 10$^4$--10$^5$ Pa [\textit{Wahr}, 2009]. However, the strength of the ice may be reduced to approximately 10$^4$ Pa [e.g., \textit{Dempsey et al.}, 1999; \textit{Lee et al.}, 2005]. Additionally, stress produced by nonsynchronous rotation is on the order of a few megapascals [\textit{Wahr}, 2009]. Thus, a crack can be generated at the bottom of the ice.

In the second mechanism, the Europan ice contains components with low eutectic points, such as sulfuric acid or ammonia. In this case, the temperature of the ice around the interface is reduced to 180--200 K [\textit{Marison et al.}, 2002; \textit{Quick and Marsh}, 2016]. Thus the brine interface is located in or bottom of the stagnant lid (Fig.~\ref{model}(c)). Ice behaves as an elastic material at temperatures below 185 K [\textit{Nimmo et al.}, 2002]. Thus, in addition to cracks forming at the bottom of the ice, cracks generated by volume expansion [\textit{Manga and Wang}, 2007] may induce deflection if they are initiated slightly under the brine interface. One advantage of crack generation by volume expansion is that a sufficient magnitude of stress can be induced [\textit{Manga and Wang}, 2007], which does not require the reduction of the strength of the ice. 

The third hypothesized mechanism involves a vertical dyke propagating to a depth of up to 1--2 km and water warming the surrounding ice to the eutectic point. Although cracks formed by volume expansion can propagate to the surface of Europa [\textit{Manga and Wang}, 2007], at a depth of approximately 1--2 km, large stress is induced by thermal contraction [\textit{Nimmo}, 2004]. If this stress prevents the vertical crack from propagating to the surface, water fills the whole crack, which warms the surrounding ice to its eutectic point, thereby forming the brine interface is at the top of the vertical dyke (Fig.~\ref{model}(d)). Although lateral stress produced by volume expansion is relaxed within one hour [\textit{Manga and Wang}, 2007], tidal stress may generate a horizontal crack at the shallow interface.

\begin{figure}[t]
\centering
\includegraphics[width=13cm] {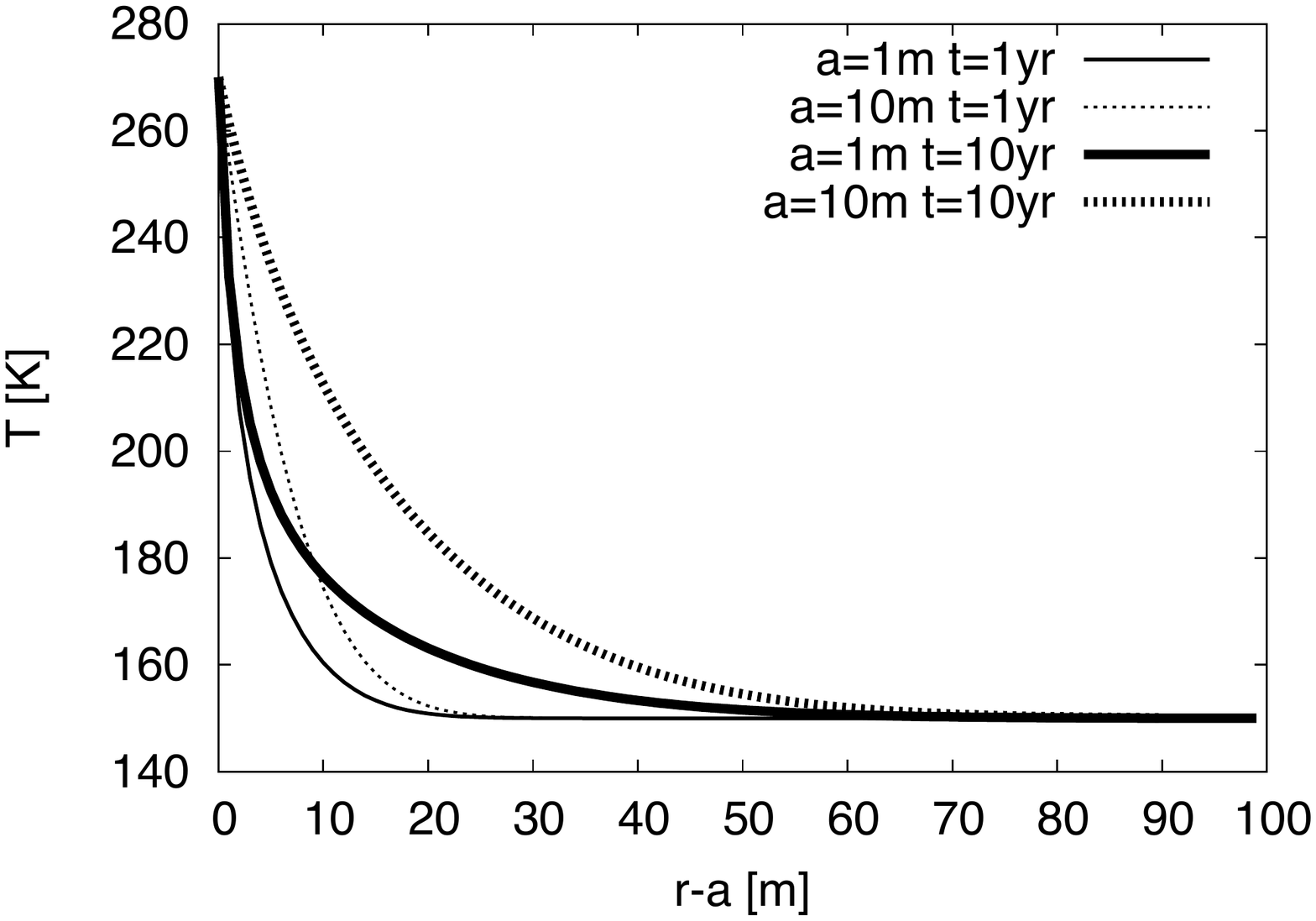}
\caption{Temperature distribution with different dyke radii and times. The temperatures of the water-filled dyke wall and initial surrounding ice were assumed to be 270 and 150 K, respectively.}
\label{temp_cond}
\end{figure}

Under the assumption that the vertical water-filled crack is cylindrical, the temperature distribution $T(r)$ of the surrounding ice is given by [\textit{Jaeger and Chamalaun}, 1966]

\begin{equation}
\frac{T(r)-T_1}{T_0-T_1}\sim1-\sqrt{\frac{a}{r}}\mathrm{erfc}\left(\frac{r-a}{2\sqrt{\kappa t}}\right),
\label{conduction}
\end{equation} 
where $T_0$, is the initial temperature of the surrounding ice, $T_1$ is the temperature of the dyke wall, $r$ is the distance from the center of the dyke, $a$ is the radius of the dyke, $\kappa$ is the thermal diffusion, $t$ is time, and efrc() is the Gauss error function, respectively. Fig.~\ref{temp_cond} shows temperature distributions under different dyke radii and at different times. $T_0$, $T_1$,  and $\kappa$ were assumed to be 150 K, 270 K, and 1.4$\times$10$^{-6}$ m$^2$ s$^{-1}$, respectively. If the radius of the dyke is 1 m, the temperature only a few meters from the water-filled dyke wall is increased to approximately 200 K. Although a dyke with $a=10$ m can warm the surrounding ice up to a distance of approximately 10 m to 200 K for 10 yr, this distance is much smaller than the radius of a typical lenticulae. Thus, it would be difficult to induce a sill via crack deflection by the warming of the ice. One possibility is that, once a horizontal crack is induced and reaches a length of a few meters, water intrudes into the horizontal crack, and the peeling of two layers occurs [\textit{Lister et al.}, 2013], which may cause further warming of the interface by the water. However, it is uncertain whether this peeling can result in a sill with a radius of a few kilometers. Thus, this hypothesis is less probable than the first and the second mechanisms. For a sill to form at shallow depth, other mechanisms, such as surface compressive stress [e.g., \textit{Menand et al.}, 2010], are more reasonable. Such mechanisms will be explored in future work. 

\section{Conclusion}
To link vertical dyke formation and the evolution of sills, the deflection of vertical cracks at the interface between two ice layers of different stiffness was considered. Whether the crack deflects or penetrates the  interface depends on the toughness of the ice [\textit{He and Hutchinson}, 1989; \textit{He et al.}, 1994]. Thus, in addition to the difference in the Young's moduli of the two regions, the occurrence of deflection was evaluated with the critical stress intensity factor of the ice taken into consideration. 

For deflection to occur, $(K_{\mathrm{I}c})_i/(K_{\mathrm{I}c})_{1}$ must be at most 0.45--0.5. Of several characteristics of the ice (porosity, temperature, grain size, and brine) that affect the critical stress intensity factor, the presence of brine in the ice can reduce $(K_{\mathrm{I}c})_i/(K_{\mathrm{I}c})_{1}$ sufficiently. On the basis of experimental results for brine-containing ice, if the brine volume fraction is more than 30 ppt, the crack may deflect along the interface. Although the chemical composition of Europan ice can be significantly different from that of terrestrial sea ice, a brine volume fraction of 30--40 ppt can be induced by an ice salinity of 1-2 percent. 

Brine is contained in ice when the ice temperature is greater than approximately 240 K, as demonstrated in numerical simulations based on reasonable estimates for the composition of eutectic ice in Europa [e.g., \textit{Zolotov et al.}, 2002].  An interface at this temperature would be too warm to induce vertical cracks by volume expansion. Thus, the generation of cracks at the bottom of the convective layer by tidal stress [\textit{Crawford and Stevenson}, 1988] and reduced strength [\textit{Dempsey et al.}, 1999] is consistent with the deflection at a brine interface with this temperature. If the ice contains components with a low eutectic point, such as ammonia or sulfuric acid [e.g., \textit{Marison}, 2004], a brine interface may be generated in the stagnant lid. Because the critical temperature for elastic behavior is approximately 185 K [\textit{Nimmo et al.}, 2002], stress generated by volume expansion can induce crack deflection at the interface. The final mechanism considered herein of warming by a water-filled dyke seems insufficient to induce an interface of a few kilometers in length.

Although many experiments on ice fracture have been conducted, insufficient data on mode II cracking and cracking at interfaces have been reported. Thus, $(K_{\mathrm{I}c})_i$ was approximated as the $K_{\mathrm{I}c}$ value of brine-containing ice when the ratio of the critical stress intensity factor was compared with $(K_d/K_p)_{\mathrm{def}}$. Because one of the two layers at the interface do not contain brine, this approximation may be oversimplified. In future work, cracking along the interface must be experimentally considered. In addition, the present experimental results on critical intensity factor are scattered because the critical stress intensity factor changes based on the conditions and specimens used in each experiment. Additionally, the chemical composition of Europan ice can be different from that of terrestrial sea ice. Thus, more studies are necessary to evaluate the deflection of vertical cracks considering the conditions of Europan ice. To assess the detailed dynamics of crack deflection, numerical simulation like that conduced by \textit{Craft et al.} [2016] is also required in future work. However, in comparison with the formation of vertical cracks in Europa, the deflection of these cracks has not thoroughly investigated. The detailed formation process of lenticulae after this deflection has been evaluated [\textit{Michaut and Manga}, 2014; \textit{Manga and Michaut}, 2017]. Thus, the deflection condition assessed in this work can serve as a first step linking vertical dykes and lenticulae formation.

\section*{Acknowledgements}
This work was supported by a JSPS Research Fellowship.

\newpage
\begin{figure}[t]
\centering
\includegraphics[width=17cm] {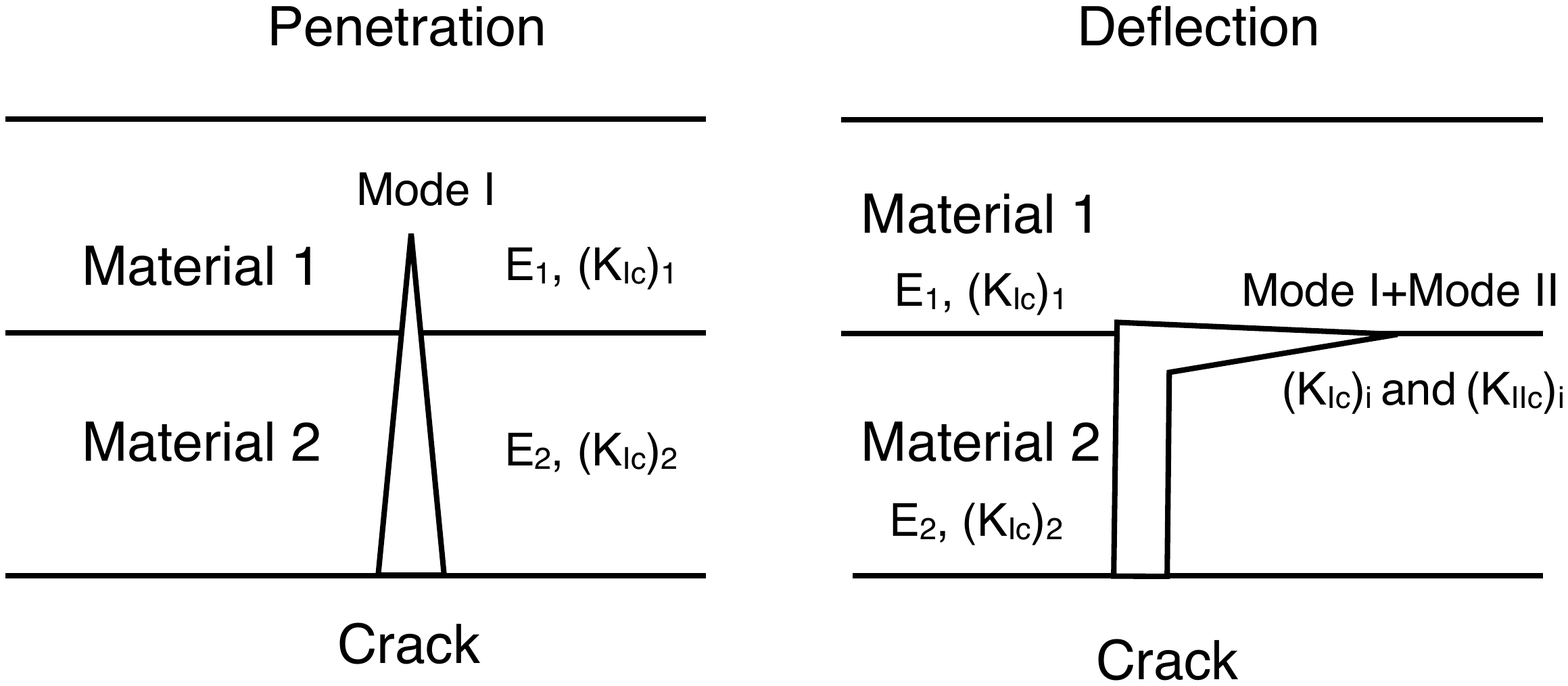}
\caption{Schematic views of the penetration and deflection of a crack at an interface. The crack propagates from material 2 to material 1. While propagation in material 1 is caused by mode I fracture, the crack propagates along the interface by a mix of mode I and II fracture.}
\label{pene_def}
\end{figure}

\newpage

\newpage

\newpage

\newpage

\end{document}